\documentstyle[epsfig,12pt]{article}  
\newcommand{\beq}{\begin{equation}}
\newcommand{\eeq}{\end{equation}}
\newcommand{\bea}{\vspace{0.25cm}\begin{eqnarray}}
\newcommand{\eea}{\end{eqnarray}}

\newcommand{\gb}{\mbox{{\boldmath
$\gamma$}}}
\newcommand{\ab}{\mbox{{\boldmath
$\alpha$}}}
\newcommand{\r}{\mbox{{\boldmath
$\rho$}}}

\newcommand{\qb}{\mbox{{\bf
q}}}
\newcommand{\pb}{\mbox{{\bf
p}}}

\newcommand{\kb}{\mbox{{\bf
k}}}
\def\lsim{\mathrel{\rlap{\lower4pt\hbox{\hskip1pt$\sim$}}
    \raise1pt\hbox{$<$}}}         %less than or approx. symbol
\def\gsim{\mathrel{\rlap{\lower4pt\hbox{\hskip1pt$\sim$}}
    \raise1pt\hbox{$>$}}}         %greater than or approx. symbol
\begin{document}
\vspace*{-2cm}
 
\bigskip

\begin{center}

  {\Large\bf
The role of finite kinematic bounds in the induced
gluon emission from fast quarks
in a finite size quark-gluon plasma
 \\
\vspace{.7cm}
  }
\medskip
{\large
  B.G.~Zakharov
  \bigskip
  \\
  }
{\it  
 L.D.~Landau Institute for Theoretical Physics,
        GSP-1, 117940,\\ Kosygina Str. 2, 117334 Moscow, Russia
\vspace{1.7cm}\\}

  {\bf
  Abstract}
\end{center}
{
\baselineskip=9pt
We study the influence of finite kinematic boundaries on the induced 
gluon radiation from a fast quark in a finite size quark-gluon plasma.
The calculations are carried out for fixed and running coupling constant.
We find that for running coupling constant the kinematic correction to the 
radiative energy loss is small for quark energy  $\gsim 5$ GeV. 
Our results differ both analytically and numerically from
that obtained by the GLV group \cite{GLV1}. 
The effect of the kinematic cut-offs is considerably smaller than 
reported in \cite{GLV1}.
\vspace{.5cm}
\\
}
%\pagebreak
%-------------------------------------------------------------
\noindent{\bf 1.} It is very likely that parton energy loss due to 
the induced gluon radiation caused by multiple scattering 
\cite{GW,BDMPS,Z1,Z2,BDMS1,GLV1,W1}
plays a major role in suppression of high-$p_{T}$ hadron
spectra in heavy-ion collisions observed at RHIC \cite{PHENIX,STAR}.
There is an attractive idea \cite{GLV2} to use this phenomenon 
(usually called 
``jet-quenching'') to obtain information
about the density of hot quark-gluon plasma (QGP) produced in 
$AA$-collisions. 
Such a jet tomographic analysis requires 
accurate methods for evaluating the induced gluon emission.
In recent years this problem has been attacked from several directions. 
In \cite{Z1} (see also \cite{Z_YAF,Z3,BSZ}) we have developed a light-cone
path integral (LCPI) approach to the induced radiation. 
The induced gluon spectrum 
was expressed through a solution of a two-dimensional Schr\"odinger
equation in the impact parameter space with an imaginary potential. 
This approach accounts for the Landau-Pomeranchuk-Migdal (LPM)
effect \cite{LP,Migdal},
finite-size and mass effects 
which are important for the QGP produced in $AA$-collisions.
In \cite{BDMPS,BDMS1} the radiative energy loss was addressed using  
diagrammatic formalism.
Similarly to the LCPI approach the BDMPS approach \cite{BDMPS,BDMS1} 
expresses the gluon spectrum
through a solution of a two-dimensional Schr\"odinger
equation in the impact parameter space.
However, the BDMPS formalism applies only in the limit of strong LPM 
suppression. In this regime it is equivalent \cite{BDMS1,BSZ} to the 
approach \cite{Z1}.
The GLV group \cite{GLV1} has developed within the soft gluon approximation
an approach in momentum space which applies to thin plasmas when 
the mean number, 
$\bar{N}$, of jet scatterings is small, and performed 
 calculations accounting for the $N=1,2,3$ rescatterings.

For applications of the formalisms \cite{Z1,BDMS1,GLV1} to the tomographic 
analysis of experimental data on $AA$-collisions 
it is important to understand the limits of applicability
of these approaches.
In the analyses \cite{Z1,BDMS1,GLV1} the QGP is modeled by a system of 
Debye screened color centers \cite{GW}, and parton scattering is treated
in the small-angle approximation.
The LCPI \cite{Z1} and BDMPS \cite{BDMPS,BDMS1} approaches, formulated in the 
impact parameter space, 
in addition, imply that the integration over 
the transverse momenta can be extended up to infinity
ignoring finite kinematic boundaries. 
The small-angle approximation for fast partons moving through the medium
should work well for parton energy $E\gg \mu_{D}$, where $\mu_{D}$ is
the Debye screening mass which plays the role of a natural infrared 
cut-off and energy scale for parton scatterings in the QGP. 
For RHIC and LHC conditions, where $\mu_{D}\sim 0.5$ GeV \cite{LH},
it means $E\gsim 3\div 5$ GeV.
It is however not clear whether the approximation of static color 
centers neglecting recoil effects is adequate for $E\sim 5$ GeV.
For such energies the kinematic constraints on the momentum
transfer $q\lsim q_{max}\sim \sqrt{E E_{th}}$, where $E_{th}\approx 3T$
is the typical thermal energy of quarks and gluons in the QGP,
may be important.
The GLV group \cite{GLV1} has reported that the kinematic
cut-offs suppress greatly the parton energy loss, $\Delta E$.
Even at $E\sim 1000\mu_{D}$ 
for the leading $N\!=\!1$ contribution to $\Delta E$ 
for a homogeneous QGP with thickness $L\approx 5$ fm
the authors have found the suppression $\sim 0.5$, and for $E\sim 10\mu_{D}$
they give the suppression $\sim 0.16$.

The approaches \cite{Z1,BDMS1} become inapplicable when the kinematic 
bounds become important.
The GLV formalism \cite{GLV1}, which does not treat accurately parton 
scattering near the kinematic limit, also cannot be used for 
quantitative calculations in this regime.
Thus, if the kinematic effect were as strong as found in \cite{GLV1},
the available approaches to the induced radiation  would be 
inapplicable even at LHC energies. 
For this reason the kinematic effect merits further
investigations. In particular, it is clearly desirable to study
the effect of the running coupling constant. The
decrease of the coupling constant near the 
kinematic bounds should act as a natural cut-off of large
parton transverse momenta and diminish the role of the recoil effects.
Another remaining open question is related to the
different cut-offs for the initial and final partons.
The authors of \cite{GLV1} have used for scattering of the radiated gluon 
the same cut-off in the momentum transfer as that for the initial parton.
However, for soft
gluons with $x\ll 1$ (hereafter $x$ is the gluon fractional momentum)
the $q_{max}$ is considerably smaller than for the initial parton.
In the present paper we address the role of the kinematic 
cut-offs accounting for the running coupling constant and different
cut-offs for the initial and final partons.
The analysis is performed for the $N\!=\!1$ scattering which
dominates the induced spectrum for RHIC and LHC conditions.
We find that although the difference in the initial and final 
state cut-offs changes the analytical form of the induced spectrum, 
numerically the effect is insignificant. 
For fixed coupling constant the kinematic corrections 
become important for $E\lsim 10\div 20$ GeV,
and for running one the kinematic 
effect is small even at $E\sim 5$ GeV. 
We find that the kinematic effect is considerably smaller than found in 
\cite{GLV1}.

\vspace{.2cm}
\noindent{\bf 2.} We consider a fast quark with energy $E$ produced at $z=0$ 
(we choose  the $z$-axis along the momentum of 
the quark) traversing a medium of thickness $L$, which eventually splits 
into a gluon and final quark with 
the energies $xE$ and $(1-x)E$ respectively. We assume that parton energies
are much larger than the thermal quasiparticle masses in the QGP.
The $N\!=\!1$ induced spectrum can be represented in the form
\beq
\frac{d P}{d
x}=
\int\limits_{0}^{L}\! d z\,
n(z)
\frac{d
\sigma^{BH}(x,z)}{dx}\,,
\label{eq:2}
\eeq
where $n(z)$ is the number density of 
the medium (the summation over the triplet (quark-antiquark) and octet
 (gluon) color states is implied on the right-hand side of (\ref{eq:2})),
and $\frac{d
\sigma^{BH}(x,z)}{dx}$
is the in-medium ($z$-dependent) Bethe-Heitler cross section. It
can be written as
\bea
\frac{d
\sigma^{BH}(x,z)}{dx}=
J_{bb}+J_{cc}+J_{dd}+2J_{bc}+2J_{cd}+2J_{db}\nonumber\\
+
2J_{ae}+2J_{af}+2J_{ag}+2J_{ah}
\label{eq:3}
\eea
with $J_{\alpha \beta}$ given by 
\beq
J_{\alpha \beta}=\frac{E}{(2\pi)^{5}}
\mbox{Re}\int d\qb d\pb T_{\alpha}(\qb,\pb)T_{\beta}^{*}(\qb,\pb)\,,
\label{eq:4}
\eeq
where the amplitudes $T_{\alpha}$ diagrammatically 
are shown in Fig.~1, $\qb$ and $\pb$ are the transverse momenta of 
the $t$-channel and emitted gluons, respectively. 
Note that the interference between the double-gluon exchange diagrams 
(e),~(f),~(g),~(h) and the diagram without gluon exchange (a) is important 
to ensure unitarity. 

The diagrams of Fig.~1 can be evaluated with the help of the 
ordinary perturbative formula 
\beq
T=\int_{0}^{\infty}dz'\int d\r g\bar{\psi}_{f}(\r,z')\gamma^{\mu}
A_{\mu}(\r,z')\psi_{i}(\r,z')\,,
\label{eq:5}
\eeq 
where $\r$ is the transverse coordinate,
$\psi_{i,f}(\r,z')$ are the wave functions of the initial and
final quarks, and $A_{\mu}(\r,z')$ is the wave function of the emitted gluon
(hereafter we omit the color factors and spin indices).
In (\ref{eq:5}) we do not explicitly 
indicate the dependence of the wave functions on the position
of the scattering center.
The quark wave functions using the light-cone spinor basis  
can be written as
\beq
\psi_j(\r,z')=\exp(iE_j z')\hat{U}_{j}\phi_j(\r,z')\,,
\label{eq:6}
\eeq
where the operator $\hat{U}_j$ reads 
\beq
\hat{U}_j=\left(1+\frac{\ab\pb+\beta m_q}{2E_j}
\right)\chi_j\,.
\label{eq:7}
\eeq
Here $\chi_j$ is the quark spinor (normalized to unity), 
$\ab=\gamma^0\gb$, $\beta=\gamma^0$, and $\pb=-i\nabla_{\perp}$.
The transverse wave function $\phi_{j}(\r,z')$ entering (\ref{eq:6}) 
is governed
by the two-dimensional Schr\"odinger equation in which $z'$ plays the role of
time
\beq
i\frac{\partial\phi_{j}(\r,z')}{\partial z'}=\left[
\frac{({\pb}^{2}+m^{2}_{q})}{2 E_{j}}+v(\r,z')\right]\,\phi_{j}(\r,z')\,,
\label{eq:7}
\eeq
where 
\beq
v(\r,z')=\delta(z'-z)
\int \frac{d\qb}{(2\pi)^{2}} \exp{(i\qb\r)}v(\qb)\,,\,\,\,\,\,\,\,\,\,\,
v(\qb)=\frac{4\pi \alpha_{s}(q)}{\qb^{2}+\mu_{D}^{2}}
\label{eq:8}
\eeq
is the potential generated by the one-gluon exchange 
between quark and the Debye screened color center.
In the longitudinal direction we treat the potential as 
a point-like.
In the same form one can represent the gluon wave function
(up to an obvious change of the spin operator and color factors).

The amplitudes entering (\ref{eq:4}) can be easily obtained 
from (\ref{eq:5}) treating in (\ref{eq:7}) the potential
$v$ as a perturbation. For the diagrams with gluon exchanges
in the $z'$ regions $0<z'<z$ and $z'>z$ the transverse wave functions
age given by the plane waves (with different transverse momenta in these
two regions of $z'$)  
\beq
\phi_j(\r,z')\propto \exp\left\{i\left[\pb_j\r-z' 
\frac{(\pb_{j}^{2}+m_{j}^{2})}{2E_j}
\right]\right\}\,.
\label{eq:9}
\eeq
The color center acts as a kick which changes the quark (or gluon) 
transverse momentum at $z'=z$. The corresponding amplitude 
$\propto v(\qb)$ for one-gluon exchange diagrams, and 
$\propto \int d\pb v(\pb)v(\qb-\pb)$ for the double-gluon exchange ones.
Note that, eventually, the $\r$-integration in (\ref{eq:5}) ensures
conservation of the transverse momentum.

To account for the kinematic boundaries we introduce in the amplitudes
the cut-off factors. 
For each $t$-channel gluon we modify the propagator introducing
the cut-off factor $\theta(q_{max}-q)$. 
Here $q_{max}$
is the upper kinematic bound on the momentum transfer
for the parton to which the $t$-channel gluon is attached.
Also, we modify the $qqg$-vertex for splitting
the initial fast quark into quark-gluon system introducing
the cut-off factor $\theta(k_{max}-k)$, $k$ is
the transverse momentum of the gluon in the frame where the 
total transverse momentum of the quark-gluon state equals zero,
and $k_{max}=E\,\mbox{min}(x,1-x)$ (here $E$ is the initial quark
energy).
The above prescription ensures that parton scattering angles are small, and
the momentum transfer does not exceed the kinematic bounds.

Using (\ref{eq:3})--(\ref{eq:8}) after straightforward but a bit cumbersome 
calculations 
the effective Bethe-Heitler cross section can be represented in 
the form
\beq
\frac{d\sigma^{BH}(x,z)}{dx}=
\frac{d\sigma_{1}^{BH}(x,z)}{dx}+
\frac{d\sigma_{2}^{BH}(x,z)}{dx}\,,
\label{eq:10}
\eeq
where 
\bea
\frac{d\sigma_{1}^{BH}(x,z)}{dx}=
\frac{2C_{T}}{\pi^{2}x}\left(1-x+\frac{x^2}{2}\right)\cdot
\int d\qb d\kb \frac{\alpha_{s}^{2}(q)}{(\qb^{2}+\mu_{D}^{2})^{2}}
\left[\theta(q_{3}-q)F(\kb,\qb,z)
\right. 
\nonumber\\
\left. 
+\theta(q_{1}-q)F(\kb,\qb (1-x),z)-
\frac{1}{N_{c}^{2}}\theta(q_{2}-q)F(\kb,\qb x,z)\right.]\,,
\label{eq:11}
\eea
\beq
F(\kb,\qb,z)=\left[\frac{\kb^{2}\Theta^{2}(\kb)}
{(\kb^{2}+\epsilon^{2})^{2}}-
\frac{(\kb-\qb)\kb \Theta(\kb)\Theta(\kb-\qb)}
{(\kb^{2}+\epsilon^{2})((\kb-\qb)^{2}+\epsilon^{2})
}\right]\cdot\left[1-
\cos\left(\frac{iz}{l(\kb,x)}
%\exp\left(\frac{i(\kb^{2}+\epsilon^{2})z}{2Ex(1-x)}
\right)
\right]\,,
\label{eq:12}
\eeq
\bea
\frac{d\sigma_{2}^{BH}(x,z)}{dx}=
\frac{2C_{T}}{C_{A}\pi^{2}x}\left(1-x+\frac{x^2}{2}\right)\cdot
\int d\qb \frac{\alpha_{s}^{2}(q)}{(\qb^{2}+\mu_{D}^{2})^{2}}
[C_{F}(\theta(q_{0}-q)-\theta(q_{2}-q))\nonumber\\
+
C_{A}(\theta(q_{2}-q)-\theta(q_{3}-q))]
\cdot
\int d\kb 
\frac{\kb^{2}\Theta^{2}(\kb)}{(\kb^{2}+\epsilon^{2})^{2}}
\left[1-\cos\left(\frac{iz}{l(\kb,x)}\right)
\right]\,
\label{eq:13}
\eea
with the following shorthands:
\beq
l(\kb,x)=\frac{2Ex(1-x)}{\kb^{2}+\epsilon^{2}}\,,
\label{eq:15}
\eeq
\beq
q_{0}=q_{max}(E),\qquad 
q_{1}=q_{max}(Ex),\qquad
q_{2}=q_{max}(E(1-x)),\qquad
q_{3}=\mbox{min}(q_{1},q_{2}),
\label{eq:13p}
\eeq 
$\Theta(\kb)=\sqrt{\alpha_{s}(k)}\theta(k_{max}-k)$,
$\epsilon^{2}=m_{q}^{2}x^{2}+m_{g}^{2}(1-x)$, $m_{q,g}$ are the thermal   
quark and gluon quasiparticle masses, $C_{T,F,A}$
%, $C_{F}$, and $C_{A}$
are the color Casimir factors of the color center, quark and gluon
respectively. Eq.~(\ref{eq:13p}) corresponds to the above described
scheme when each scattered parton has its own $q$-cut-off
factor. 
Note that in the soft gluon limit $x\ll 1$ our formulas
do not reduce to that of Ref.~\cite{GLV1}.
If one uses for the final partons the same $q_{max}$ as 
for the initial quark as was done in \cite{GLV1},
the second term on the right-hand side of (\ref{eq:10}) vanishes.
This term emerges inevitably because the initial and final partons have
different phase space for their scattering.  
Below for comparison with \cite{GLV1} we also present the results 
for $q_{i}=q_{max}(E)$ as in \cite{GLV1}.

The quantity $L_{f}=l(\kb=0,x)$ characterizes 
the longitudinal scale of gluon emission, i.e.,
the gluon formation length. 
The induced spectrum depends crucially on the ratio $L_{f}/L$
\cite{Z1,Z2,Z5}. For gluons with small formation 
length $L_{f}\ll L$ the finite-size effects 
caused by the oscillating cosine on the right-hand side of 
(\ref{eq:12}),~(\ref{eq:13})
becomes small. In this regime the rapidly oscillating cosine
can be neglected, and the effective cross section (\ref{eq:10}) 
becomes equal to 
the ordinary Bethe-Heitler one, i.e., to the cross section for a quark 
which approaches the color center from
outside. 
In contrast, when $L_{f}\gsim L$ the finite size effects
due to the cosine in (\ref{eq:12}),~(\ref{eq:13}) suppress greatly the 
radiation rate as compared to the Bethe-Heitler one \cite{Z2,Z5}.
This suppression, physically, is connected with small transverse size
of the $qg$ system (it is $\propto L$). In this regime the $t$-channel 
gluons cannot distinguish the $|q\rangle$ and $|qg\rangle$ Fock components
of the physical quark and for this reason the gluon emission turns out 
to be suppressed. 
One remark regarding the Bethe-Heitler regime 
for $L_{f}\ll L$ is in order here.
Diagrammatically, the ordinary Bethe-Heitler cross section is given
by the diagrams (b), (c), (d) of Fig.~1 involving only one-gluon exchange.
However, our formulas include the interference between
the diagram (a) and (e),~(f),~(g),~(h). The explanation of this fact is
as follows. For a quark incident on the color center from outside
the amplitudes (b),~(c),~(d) should be evaluated integrating over $z'$ 
in (\ref{eq:5}) from $-\infty$ (with usual adiabatic switching off of 
the coupling
constant for $|z'|\to \infty$). For a quark produced in
a hard reaction at $z'=0$ the $z'$-integration region is $(0,\infty)$.
This gives rise to additional endpoint terms 
(corresponding to $z'=0$) in the cross section which are
absent when the lower limit equals $-\infty$.
However, similar endpoint terms emerge for the interference term
involving the double-gluon exchange diagrams
as well. They cancel exactly the endpoint terms stemming from the 
graph (b),~(c),~(d). As a result for $z\to \infty$ our effective
cross section (\ref{eq:10}) equals the ordinary Bethe-Heitler one.

It should be noted that our method (and any other one based 
on the GW model \cite{GW} for the QGP) 
can only give an estimate for the kinematic correction. It is inapplicable 
in the regime of strong kinematic suppression
when the spectrum becomes very sensitive
to the detailed form of the kinematic cut-offs. This fact is closely
connected with the anti-leading log character of the $\qb,\kb$-integrations
in (\ref{eq:11}). Contrary to the ordinary leading log situation, say,
in $\gamma^{*}\to q\bar{q}$ transition in deep inelastic 
scattering, where the typical values of the momentum transfer $q$ is 
smaller than the internal momentum $k$, in the case of the induced  
gluon emission in the high energy limit when $L_{f}\gg L$ the dominating
contribution to the induced spectrum comes from $q\gsim k$. 
A detailed discussion of this phenomenon
is given in \cite{Z5}. 

The effective Bethe-Heitler cross section  
evaluated without kinematic cut-offs, i.e.,
with $q_{i}=k_{max}=\infty$, is given by the first term on the right-hand
side of (\ref{eq:10}). 
After the Fourier transform it can be represented in the impact parameter
space in the form obtained previously \cite{Z4,Z5} within the LCPI formalism 
\beq
\frac{d
\sigma^{BH}(x,z)}{dx}=\mbox{Re}
\int d\r\,
\Psi^{*}(\r,x){\sigma}_{3}(\rho,x)\Psi_{m}(\r,x,z)\,,
\label{eq:16}
\eeq
where $\Psi(\r,x)$ is the ordinary
light-cone wave function for the $q\to g q$ transition in vacuum, 
$\Psi_{m}(\r,x,z)$ is the $z$-dependent light-cone wave function describing the 
quark-gluon Fock component at the longitudinal coordinate $z$, and 
${\sigma}_{3}(\rho,x)$ is the three-body cross section of 
a $q\bar{q}g$ system with a particle in the medium
(the explicit form of the wave functions and three-body cross section
can be found in \cite{Z5}).
In the $q\bar{q}g$ system antiquark is located (in the transverse space)
in the center of mass of the $qg$ pair, and the relative separations
satisfies the relation 
$(\r_{g}-\r_{\bar{q}})x=(1-x)(\r_{\bar{q}}-\r_{q})$.

\vspace{.2cm}
\noindent{\bf 3}. We have performed numerical calculations for fixed and 
running coupling constant. In the first case we take 
$\alpha_{s}=0.5$ \cite{BDMS2}.
For running coupling constant we use the one-loop formula with 
$\Lambda_{QCD}=0.3$ GeV frozen at the value $\alpha_{s}=0.65$.
In this case $\alpha_{s}$ approximately satisfies the relation
\beq
\int_{\mbox{0}}^{\mbox{\small 2 GeV}}\!dk\frac{\alpha_{s}(k)}{\pi}
\approx 0.36 \,\,
\mbox{GeV}\,
\label{eq:17}
\eeq  
obtained from the analysis of the heavy quark energy loss \cite{DKT}.
We have carried out the calculations for expanding plasma. 
We use the Bjorken \cite{Bjorken} model with
$T\tau^{3}=T_{0}\tau^{3}_{0}$, and take the initial conditions
suggested in \cite{FMS} for heavy ion central collisions at RHIC:
$T_{0}=446$ MeV and $\tau_{0}=0.147$ fm.
For the upper limit of the $z$-integration in (\ref{eq:2}) we take 
$L=R_{A}\approx 6$ fm.
For quark and gluon quasiparticle masses
we use the values obtained in \cite{LH} from the lattice data
$m_{q}\approx 0.3$ and $m_{g}\approx 0.4$ GeV.
With the above value of $m_{g}$ from the perturbative relation 
$\mu_{D}=\sqrt{2}m_{g}$ one obtains for the Debye screening mass
$\mu_{D}\approx 0.57$ GeV. 
For the energy dependence of the maximum momentum transfer we take 
$q_{max}^{2}(E)\approx E \bar{E}_{th}$ with
$\bar{E}_{th}=750$ MeV. It is smaller than 
$q_{max}^{2}(E)\approx 3E\mu_{D} $ used in \cite{GLV1}.

In Figs.~2,~3 we plot the induced gluon spectrum  
for $E=$5, 10 and 20 GeV
evaluated using 
(\ref{eq:2}),~(\ref{eq:10})--(\ref{eq:13p}) 
with (solid line) and without (dashed line) the kinematic cut-offs
for fixed and running $\alpha_{s}$. 
The results with kinematic cut-offs have been obtained for $q_{i}$ given in 
(\ref{eq:13p}). For this version we also plot  
the spectrum without the second term in (\ref{eq:10}) (dotted line).
In Figs.~2,~3 we also show the 
results for the same kinematic cut-offs for initial and final partons
 obtained with $q_{i}=q_{max}(E)$ (long-dashed line). 
From Figs.~2,~3 one sees that the kinematic cut-offs become especially 
important 
when the energy of the radiated gluon (or the final quark) $\lsim 1\div 2$ GeV.
The kinematic correction is smaller for running coupling constant.
It is also seen that for fixed coupling constant the relative contribution
from the second term in (\ref{eq:10}) is larger.
It is natural since for the running $\alpha_{s}$ the contribution
from large transverse momenta is suppressed.
The total spectrum in the above two scheme of the $q$-cut-off turns out to be 
approximately the same.

To illustrate the effect of the kinematic cut-offs on the quark energy
loss in Fig.~4 we plot the energy dependence of the kinematic $K$-factor
\beq
K(E)=\frac{\Delta E_{f.b.}}{\Delta E_{i.b.}}\,,
\label{eq:18}
\eeq
where $\Delta E_{f.b.}$ and $\Delta E_{i.b.}$ are the quark
energy losses evaluated with (for $q_{i}$ as given in (\ref{eq:13p}))
and without kinematic constraints, 
respectively. We define $\Delta E$ as
\beq
\Delta E=E\int_{x_{min}}^{x_{max}}dx x \frac{dP}{dx}
\label{eq:19}
\eeq
with $x_{min}=m_{g}/E$, $x_{max}=1-m_{q}/E$.
Fig.~4 demonstrates that for fixed coupling constant the kinematic 
cut-offs are important for $E\lsim 20$ GeV. For running coupling
the kinematic correction is small even for $E\sim 5$ GeV. One can see that 
the difference between the cut-offs given in (\ref{eq:13p}) (thick curves) and 
$q_{i}=q_{max}(E)$ (thin curves) is less than $\sim 10\%$.

The kinematic $K$-factor shown in Fig.~4 is obtained for an expanding plasma
with $n(z)\propto 1/z$. For comparison with the analysis \cite{GLV1}
we have also carried out calculations for a homogeneous plasma.
In this case  the kinematic effect is weaker
since the relative contribution of the region of small $z$ where
the typical parton transverse momenta are large 
(they are $\propto 1/\sqrt{z}$) is smaller. 
For correspondence with \cite{GLV1} we have taken 
$q_{max}^{2}=3E\mu_{D}$ (the same for the initial and final partons) 
and $k_{max}^{2}=4E^{2}\min(x^{2},x(1-x))$ used in \cite{GLV1} which
give somewhat smaller kinematic effect than our cut-offs.
We have obtained a small kinematic effect, 
say, $K\approx 0.9$ at $E=5$ GeV and $K\approx 0.94$ at $E=10$ GeV.
(we have used fixed coupling constant as in \cite{GLV1}).
It is considerably larger than 
the suppression reported in \cite{GLV1}  ($\sim 1/6$ for 
$E=5$ GeV). 
In connection with strong kinematic suppression reported in \cite{GLV1}
we would like to emphasize one more time that 
within the approximation of static color centers \cite{GW} when the 
kinematic cut-offs are introduced by hand the regime of strong kinematic
effect cannot be described accurately.
It is clear that the analysis of the induced radiation
in this regime requires an accurate treatment of the recoil effects.
In this case the fast partons moving through QGP and partons from 
QGP should be treated on an even footing. Note also that in this 
regime suppression of the radiative energy loss may largely be compensated 
by the collisional energy loss due to strong recoil effects.

In summary, 
the form of the induced gluon spectrum 
obtained in the present analysis shows that the kinematic effect is relatively
small and is mainly important near to the endpoints $x\sim 0$ and $x\sim 1$
when the energy of the radiated gluon (or of the final quark) is about 
$\sim 2\div 3$ units of the Debye screening mass, i.e., about $1\div 2$ 
GeV for RHIC and LHC conditions. For fixed coupling constant the kinematic 
correction to the quark energy loss becomes small for $E\gsim 20$ GeV, 
for running coupling
constant it is small even at $E\sim 5$ GeV. 
The kinematic effect 
found in our analyses is considerably smaller than reported in \cite{GLV1}.
Our results say that in the region of the gluon 
fractional momentum
$\delta \lsim x \lsim 1-\delta$ ($\delta \sim (2\div 3)\mu_{D}/E$),
the induced spectrum can be evaluated to reasonable accuracy within 
the LCPI approach \cite{Z1} which ignores the kinematic bounds.
This approach can be used for evaluation of the energy 
loss and nuclear suppression factor \cite{BDMS2} 
for RHIC and LHC energies.

\bigskip
\noindent {\large \bf Acknowledgements.}
I thank N.N.~Nikolaev for discussion of the results. 
I am grateful to the High Energy Group of 
the ICTP for the kind hospitality during my
visit to Trieste where the present calculations were carried out.

\newpage

%------------------------------------------------------------------
\begin{center}
{\Large \bf Figures}
\end{center}
%------------------------------------------------------------------

\begin{figure}[h]
\begin{center}
\epsfig{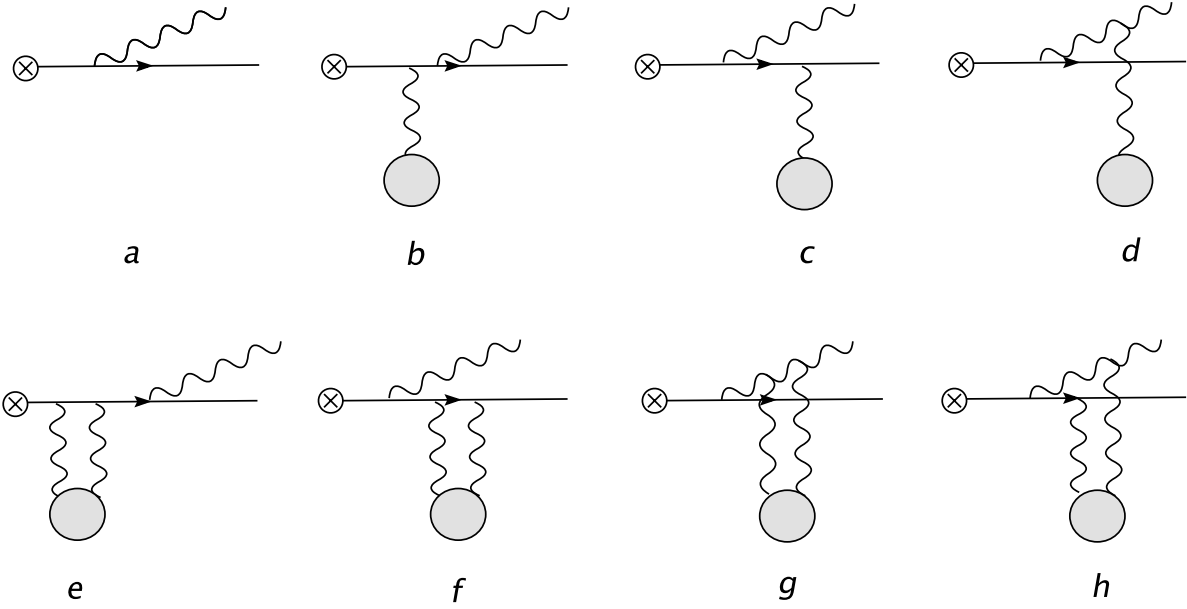}
\end{center}
\caption[.]{
The set of the Feynman diagrams for the 
$N\!=\!1$ contribution to the induced gluon spectrum.
}
\end{figure}

\begin{figure}[h]
\begin{center}
\epsfig{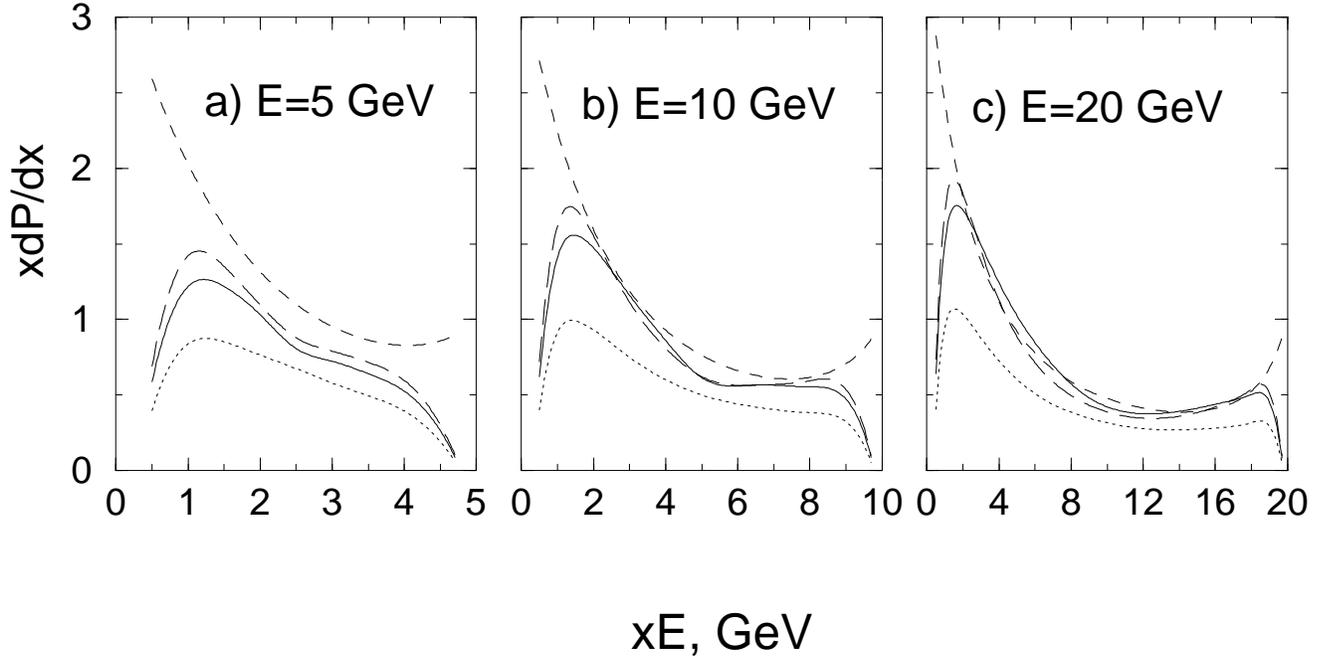}
\end{center}
\caption[.]{
The spectrum of the induced $q\to g q$ transition versus the gluon energy $xE$
for RHIC conditions for fixed coupling constant.
The solid lines are for $q_{i}$ given in (\ref{eq:13p}) and 
the long-dashed lines are for the same $q$-cut-offs for the initial and 
final partons with $q_{i}=q_{max}(E)$.
The dashed lines show the spectrum obtained without kinematics cut-offs.
The spectrum without the second term in (\ref{eq:10}) for
$q_{i}$ given in (\ref{eq:13p}) is shown by 
the dotted  curves.
}
\end{figure}

\begin{figure}[h]
\begin{center}
\epsfig{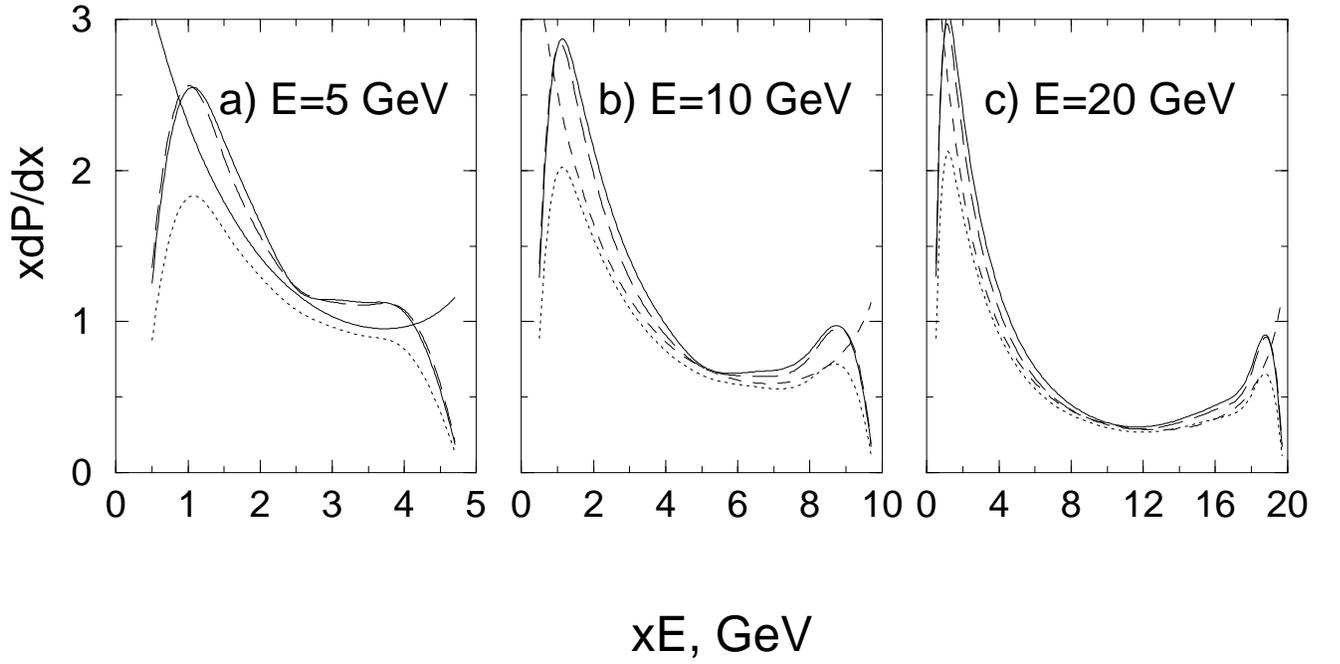}
\end{center}
\caption[.]{
The same as in Fig.~2 but for running coupling constant.
}
\end{figure}

\begin{figure}[t]
\begin{center}
\epsfig{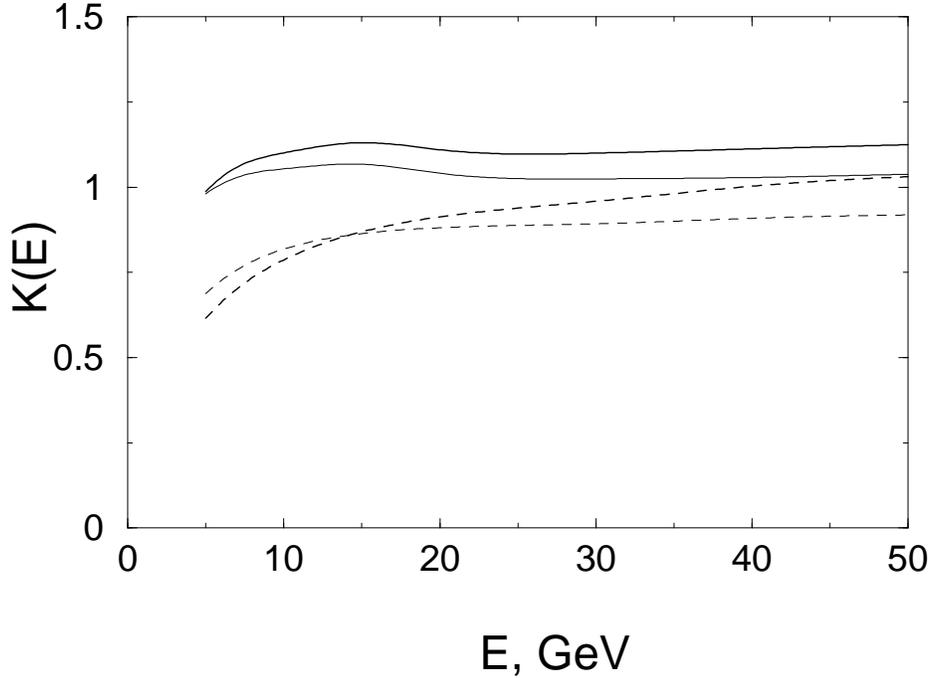}
\end{center}
\caption[.]{
The energy dependence of the kinematic 
$K$-factor (\ref{eq:18}) for RHIC conditions for running (solid line) 
and fixed (dashed line) coupling constant.
The thick lines are for the $q$-cut-off given in (\ref{eq:13p}) 
and the thin lines are for $q_{i}=q_{max}(E)$.
}
\end{figure}

\end{document}